\documentclass{article}
\usepackage{iclr2025_conference,times}
\usepackage[utf8]{inputenc}
\usepackage[T1]{fontenc}
\usepackage{hyperref}
\usepackage{url}
\usepackage{float}
\usepackage{tikz}
\usetikzlibrary{shapes,arrows,positioning,calc}
\definecolor{vennhuman}{HTML}{2A78D6}
\definecolor{vennllm}{HTML}{EB6834}
\definecolor{navytxt}{RGB}{20,20,20}
\definecolor{graytxt}{RGB}{97,97,97}
\definecolor{stgA}{HTML}{4C86C8}
\definecolor{stgB}{HTML}{3F78B8}
\definecolor{stgC}{HTML}{3369A7}
\definecolor{stgD}{HTML}{285A96}
\definecolor{stgE}{HTML}{1E4B82}
\definecolor{stgF}{HTML}{143A6B}

\iclrfinalcopy %
\title{Complementary Vulnerabilities: Auditing and Governing Clinician-LLM Error Overlap}

\author{Mohsen Amoei \\
McGill University, Research Institute of the \\
McGill University Health Centre, and Mila \\
Montreal, QC, Canada \\
\texttt{mohsen.amoei@mail.mcgill.ca} \\
\And
Kathleen Fraser \\
University of Ottawa \\
Faculty of Engineering \\
Ottawa, ON, Canada \\
\texttt{kathleen.fraser@uottawa.ca} \\
\And
Dan Poenaru \\
McGill University, Research Institute of the \\
McGill University Health Centre, and Mila \\
Montreal, QC, Canada \\
\texttt{dan.poenaru@mcgill.ca} \\
}

\begin{document}
\maketitle

\lhead{Accepted to SMASH 2026}

\begin{abstract}
Health systems are deploying large language models (LLMs) into clinical workflows, often on the assumption that a second automated check can only reduce error. But a second reasoner reduces error only insofar as it fails differently from the first. We say that a clinician and a model have \emph{complementary vulnerabilities} to the extent that they break down through different mechanisms and on different cases. The two reasoners' failure substrates differ: they are embodied and cognitive in the clinician, and statistical and generative in the model -- so their weak points need not line up. That divergence is plausible rather than guaranteed, and the independence it creates is fragile, eroding when clinicians defer to the model or the model conforms to the clinician. We therefore propose a predeployment audit in which each site measures error overlap on its own cases. We also propose a governance model with which the site keeps those errors independent over time. Health systems should then judge a clinical artificial intelligence (AI) tool by whether its failures overlap with their clinicians' failures, as well as by its accuracy.
\end{abstract}

\section{Introduction}
Diagnostic error is one of the most consequential threats to patient safety, with an estimated 795{,}000 people in the United States suffering permanent disability or death each year~\citep{r1}. One proposed remedy, cognitive forcing, failed to reduce diagnostic error in a trial of medical students~\citep{r2}. Against this baseline, LLMs are proposed for triage, differential diagnosis, and second-opinion workflows, framed as a redundant check, a second reader expected to catch what the clinician misses. But redundancy improves safety only when the second reader does not fail on the same cases for the same reasons. Guidance for clinical AI consistently calls for human oversight, but rarely specifies where to place it or how to tell whether it is working~\citep{r3,r4}. In crowd-worker experiments, human-AI teams outperformed either party alone only under specific conditions~\citep{r5}, which leaves open which conditions hold for a given deployment. We argue the deciding condition is whether the clinician and the model err on the same cases, and we translate that principle into an audit and a governance model a health system can operate.

\section{Divergent failure mechanisms}
Human diagnostic failure is embodied and state-dependent. Among 6{,}695 surveyed physicians, burnout and fatigue were independently associated with self-reported major errors~\citep{r6}, and clinicians overweight recently encountered diagnoses~\citep{r7}. Clinical LLM failure has a very different signature. Models fabricate content~\citep{r8}, their accuracy shifts when clinically identical information is reordered~\citep{r9}, they endorse unsafe premises rather than correct them~\citep{r10}, and repeated runs on identical inputs return inconsistent answers~\citep{r11}. The most treacherous pattern for deployers comprises failures that share a label across the two reasoners, such as anchoring and overconfidence. In anchoring, the clinician fixates on a first impression~\citep{r12,r13}, while an LLM anchored in a similar way in clinical decision tasks~\citep{r14}, plausibly because of prompt content and ordering. A shared label does not necessarily prove shared cases. An LLM trained largely on human-generated medical text may also absorb the very heuristics that mislead clinicians, a common-cause path to correlated error~\citep{r15}. Sites must therefore measure complementarity on the same cases.

\section{The complementary-vulnerabilities framework}\label{sec:framework}
\begin{figure}[t]
\centering
\resizebox{0.50\textwidth}{!}{%
\begin{tikzpicture}[font=\sffamily, >=latex]
\def\R{3.0}
\def\cx{1.6}
\coordinate (LC) at (-\cx,0);
\coordinate (RC) at (\cx,0);
\fill[vennhuman, fill opacity=0.10] (LC) circle [radius=\R];
\fill[vennllm, fill opacity=0.10] (RC) circle [radius=\R];
\begin{scope}
  \clip (LC) circle [radius=\R];
  \fill[black, fill opacity=0.06] (RC) circle [radius=\R];
\end{scope}
\draw[very thick, vennhuman!80!black] (LC) circle [radius=\R];
\draw[very thick, vennllm!80!black] (RC) circle [radius=\R];
\node[font=\bfseries] at (-2.8,3.4) {HUMAN};
\node[font=\bfseries] at ( 2.8,3.4) {LLM};
\node[align=center, font=\small] at (-2.75,0.05)
  {fatigue / burnout\\[2pt] affective bias\\[2pt] availability bias\\[2pt] gut feelings};
\node[align=center, font=\small] at (2.75,0.05)
  {hallucination\\[2pt] prompt fragility\\[2pt] sycophancy\\[2pt] non-reproducibility};
\node[align=center, font=\bfseries\scriptsize] at (0,1.05) {SHARED\\ LABELS\,?};
\node[align=center, font=\scriptsize] at (0,-0.2)
  {anchoring\\[1pt] premature closure\\[1pt] confirmation\\[1pt] overconfidence};
\end{tikzpicture}%
}
\caption{Illustrative map of human and LLM failure mechanisms, drawn as largely non-overlapping sets. The grouping is the authors' expository choice, neither exhaustive nor validated. We drew its items from different levels, from physiological state to model behavior, and use only the substrate difference beneath them. The shaded region holds failures that share a label but may arise from different mechanisms, so sites must measure same-case overlap locally.}
\label{fig:venn}
\end{figure}
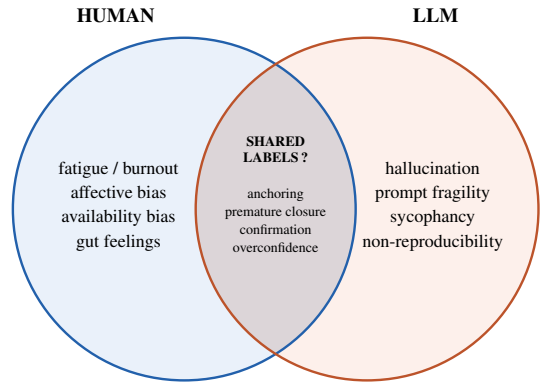

In reliability engineering and ensemble learning, two components fail redundantly when they err on the same inputs for related reasons. They fail complementarily when their errors are less correlated, so each can catch what the other misses~\citep{r16}. We say that a clinician-LLM pair has \emph{complementary vulnerabilities} to the extent that its errors fall in the second regime. The precedent for caution is quite direct. In multiversion software engineering, independently developed program versions were assumed to fail independently, yet across 27 versions each subjected to one million tests they failed together substantially more frequently than independence predicted~\citep{r17}. Such coincident failures are expected when certain inputs are intrinsically harder, so even differently designed systems still fail on the same difficult cases~\citep{r18}. The clinical counterpart is the shared-error case, where hard or ambiguous presentations may defeat both clinician and model, yet by different routes. Complementarity is therefore a falsifiable, local claim about a particular task, case mix, model version, and workflow. \citet{r19} compared physicians and LLMs on the same 2{,}133 vignettes and found that human-model error diversity exceeded model-model diversity. Independence measured in isolation can also collapse once the two interact. In one randomized trial's post hoc analysis of 58 sampled cases, the model reproduced all of the clinician's listed diagnoses in 48\% of cases when it answered second, versus 3\% when it answered first~\citep{r20}. In a meta-analysis, teams did not reliably exceed the better single agent~\citep{r21}. Other researchers have addressed parts of this problem. \citet{r22} taught an AI system when to defer to clinicians, to improve team accuracy. \citet{jain2025} treat human-AI complementarity as a way to improve human oversight of AI outputs, and measure it on annotation tasks, not on a hospital's own cases. Public health and regulatory bodies have set out principles and lifecycle stages for clinical AI~\citep{r3,r4}, and \citet{li2026} list error-complementarity matrices among their evaluation measures. None of them tells a site how to measure clinician-model error overlap before go-live, or how to base adoption on it. We propose that each site measure this overlap on its own cases before adopting an LLM. The people who govern the tool should then keep the two sets of errors independent. Figure~\ref{fig:venn} summarizes this framing.

\section{A paired-error audit}\label{sec:audit}
Evaluators build the audit on a paired, case-level record. They record the unaided clinician's diagnosis and plan before model reveal, run the locked candidate LLM on the same information, and score both against a prespecified reference standard. They then classify each case as a shared success, a model-only error (substitution risk), a clinician-only error (rescue potential), or a shared error (common-mode failure). A site that looks only at overall accuracy cannot see this breakdown. Two models can both be 90\% accurate while one fails exactly where clinicians fail and the other catches clinician misses. Sites should report the paired counts and the double-fault rate (both wrong together) with a correlation coefficient~\citep{r16}, stratified by case mix. They should then repeat the audit with clinicians using the model, because offline they can measure only rescue potential. As a sizing principle, a serious-harm outcome with a near 1\% prevalence implies a sample on the order of 1{,}000 adjudicated cases to expect 10 such events, which is a floor for detection rather than precise estimation. Before the audit, the site sets thresholds for two go/no-go decisions. The model must be accurate enough and share few failures with clinicians. Harm from model-only errors must stay manageable, and clinicians must keep their independent judgment. Where overlap concentrates in identifiable strata, the site can restrict deployment to the case mix where complementarity holds.

We propose a worked example to make the thresholds concrete. Suppose a site audits 1{,}000 adjudicated cases and records clinician-only error in 8.0\%, model-only error in 6.0\%, and shared error in 1.5\%. Clinician error is then 9.5\% overall and model error 7.5\%, so two readers failing independently would coincide on 0.71\% of cases. The observed 1.5\% (95\% Wilson confidence interval 0.9\% to 2.5\%) is 2.1 times that. Their errors are therefore positively associated, so the site cannot count on independence. The site should state its no-go threshold in those terms, for example a double-fault rate above 1.2 times its independence expectation. Overlap in high-acuity presentations matters more than the same overlap in self-limiting complaints, so sites should set a separate, tighter ceiling for serious-harm cases in advance. If the confidence interval on the overlap estimate spans the threshold, the audit is underpowered, and the site must either adjudicate more cases or restrict the deployment. At the example's error rates, a pair with independent errors passes this test in 3\% of 1{,}000-case audits and 60\% of 20{,}000-case audits. When the clinician and the model disagree, the site should send the case to an independent second clinician, because deferring to either party would erode the independence the site measured.

\section{Governance as preservation of complementarity}\label{sec:governance}
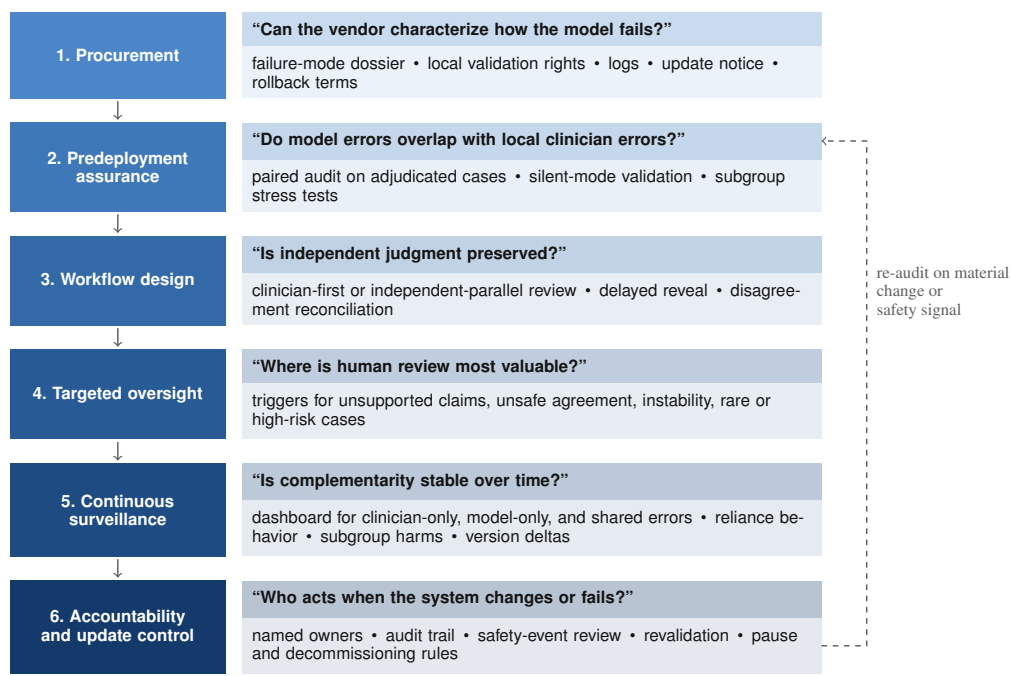
\begin{figure}[t]
\centering
\resizebox{0.95\textwidth}{!}{%
\begin{tikzpicture}[font=\sffamily, inner sep=0pt,
  qband/.style={text=navytxt, font=\sffamily\bfseries\footnotesize, inner sep=5.5pt, text width=10.5cm, align=left},
  cbody/.style={text=navytxt, font=\sffamily\footnotesize, inner sep=5.5pt, text width=10.5cm, align=left}]
\node[qband, fill=stgA!30, anchor=north west] at (0,0) (q1) {``Can the vendor characterize how the model fails?''};
\node[cbody, fill=stgA!11, anchor=north west] at (q1.south west) (b1) {failure-mode dossier \,\textbullet\, local validation rights \,\textbullet\, logs \,\textbullet\, update notice \,\textbullet\, rollback terms};
\fill[stgA] ($(q1.north west)+(-4.35,0)$) rectangle ($(b1.south west)+(-0.3,0)$);
\node[text=white, font=\sffamily\bfseries\small, align=center, text width=3.85cm] at ($($(q1.north west)+(-4.35,0)$)!0.5!($(b1.south west)+(-0.3,0)$)$) {1.~Procurement};
\node[qband, fill=stgB!30, anchor=north west] at ($(b1.south west)+(0,-4.5mm)$) (q2) {``Do model errors overlap with local clinician errors?''};
\node[cbody, fill=stgB!11, anchor=north west] at (q2.south west) (b2) {paired audit on adjudicated cases \,\textbullet\, silent-mode validation \,\textbullet\, subgroup stress tests};
\fill[stgB] ($(q2.north west)+(-4.35,0)$) rectangle ($(b2.south west)+(-0.3,0)$);
\node[text=white, font=\sffamily\bfseries\small, align=center, text width=3.85cm] at ($($(q2.north west)+(-4.35,0)$)!0.5!($(b2.south west)+(-0.3,0)$)$) {\mbox{2.~Predeployment} assurance};
\draw[->, thick, graytxt] ($(b1.south west)+(-2.325,-0.05)$) -- ($(q2.north west)+(-2.325,0.05)$);
\node[qband, fill=stgC!30, anchor=north west] at ($(b2.south west)+(0,-4.5mm)$) (q3) {``Is independent judgment preserved?''};
\node[cbody, fill=stgC!11, anchor=north west] at (q3.south west) (b3) {clinician-first or independent-parallel review \,\textbullet\, delayed reveal \,\textbullet\, disagreement reconciliation};
\fill[stgC] ($(q3.north west)+(-4.35,0)$) rectangle ($(b3.south west)+(-0.3,0)$);
\node[text=white, font=\sffamily\bfseries\small, align=center, text width=3.85cm] at ($($(q3.north west)+(-4.35,0)$)!0.5!($(b3.south west)+(-0.3,0)$)$) {3.~Workflow design};
\draw[->, thick, graytxt] ($(b2.south west)+(-2.325,-0.05)$) -- ($(q3.north west)+(-2.325,0.05)$);
\node[qband, fill=stgD!30, anchor=north west] at ($(b3.south west)+(0,-4.5mm)$) (q4) {``Where is human review most valuable?''};
\node[cbody, fill=stgD!11, anchor=north west] at (q4.south west) (b4) {triggers for unsupported claims, unsafe agreement, instability, rare or high-risk cases};
\fill[stgD] ($(q4.north west)+(-4.35,0)$) rectangle ($(b4.south west)+(-0.3,0)$);
\node[text=white, font=\sffamily\bfseries\small, align=center, text width=3.85cm] at ($($(q4.north west)+(-4.35,0)$)!0.5!($(b4.south west)+(-0.3,0)$)$) {4.~Targeted oversight};
\draw[->, thick, graytxt] ($(b3.south west)+(-2.325,-0.05)$) -- ($(q4.north west)+(-2.325,0.05)$);
\node[qband, fill=stgE!30, anchor=north west] at ($(b4.south west)+(0,-4.5mm)$) (q5) {``Is complementarity stable over time?''};
\node[cbody, fill=stgE!11, anchor=north west] at (q5.south west) (b5) {dashboard for clinician-only, model-only, and shared errors \,\textbullet\, reliance behavior \,\textbullet\, subgroup harms \,\textbullet\, version deltas};
\fill[stgE] ($(q5.north west)+(-4.35,0)$) rectangle ($(b5.south west)+(-0.3,0)$);
\node[text=white, font=\sffamily\bfseries\small, align=center, text width=3.85cm] at ($($(q5.north west)+(-4.35,0)$)!0.5!($(b5.south west)+(-0.3,0)$)$) {5.~Continuous surveillance};
\draw[->, thick, graytxt] ($(b4.south west)+(-2.325,-0.05)$) -- ($(q5.north west)+(-2.325,0.05)$);
\node[qband, fill=stgF!30, anchor=north west] at ($(b5.south west)+(0,-4.5mm)$) (q6) {``Who acts when the system changes or fails?''};
\node[cbody, fill=stgF!11, anchor=north west] at (q6.south west) (b6) {named owners \,\textbullet\, audit trail \,\textbullet\, safety-event review \,\textbullet\, revalidation \,\textbullet\, pause and decommissioning rules};
\fill[stgF] ($(q6.north west)+(-4.35,0)$) rectangle ($(b6.south west)+(-0.3,0)$);
\node[text=white, font=\sffamily\bfseries\small, align=center, text width=3.85cm] at ($($(q6.north west)+(-4.35,0)$)!0.5!($(b6.south west)+(-0.3,0)$)$) {6.~Accountability and update control};
\draw[->, thick, graytxt] ($(b5.south west)+(-2.325,-0.05)$) -- ($(q6.north west)+(-2.325,0.05)$);
\draw[->, thick, dashed, graytxt] (b6.east) -- ++(0.85,0) |- (q2.east)
  node[pos=0.35, right=3pt, align=left, font=\footnotesize, text=graytxt, inner sep=2pt] {re-audit on material\\change or\\safety signal};
\end{tikzpicture}%
}
\caption{Governance operating model for preserving clinician-model error independence across the deployment lifecycle.}
\label{fig:governance}
\end{figure}

Health agencies already call for lifecycle governance and monitoring of clinical AI~\citep{r3,r4}. We give that governance an organizing target, summarized in Figure~\ref{fig:governance}. First, automation bias is a complementarity failure. In a randomized trial, physicians given incorrect LLM recommendations lost 14 to 18 percentage points of accuracy despite 20 hours of AI-literacy training~\citep{r23}. Sites should therefore have clinicians judge first or in parallel, and treat disagreement as a safety signal. These failures are often covert, because a hallucinated reference or a confidently stated wrong answer reads much like a correct one, so reviewers should target their oversight by failure mechanism. Model-specific triggers include hallucination, unsupported citations, agreement with unsafe premises, and instability across runs. Case-specific triggers include rare, atypical, or high-acuity presentations. Reviewers should not aim to maximize agreement, since agreement may be shared success or shared failure. Independent judgment matters whichever reasoner checks the other, because a checker helps only if it can still catch the other's errors on its own. Second, sites should track \emph{complementarity drift}. A model can improve on average while becoming a worse complement, as its errors migrate into the strata where clinicians err, or as users learn to defer. Third, buyers should demand failure-mode evidence, silent-mode validation rights, and the logs required for paired-error monitoring.

We place these three elements at defined points in Figure~\ref{fig:governance}. At procurement (stage 1), buyers contract for failure-mode evidence and validation rights, and they make purchase conditional on a signed failure-mode dossier and log access. In predeployment assurance (stage 2), evaluators run the paired-error audit on the site's own adjudicated cases and make the two go/no-go decisions above. In workflow design (stage 3), the site has clinicians record their own judgment before model reveal, in every case. In targeted oversight (stage 4), reviewers check the cases flagged by the model-specific and case-specific triggers named above, instead of a fixed share of cases. In continuous surveillance (stage 5), the site recomputes the paired counts on adjudicated samples from practice. Reviewers treat a rising double-fault rate, or a fall in the clinician errors that the model catches, as complementarity drift. In accountability and update control (stage 6), named people can pause the tool, and they must order a new audit after any material version change or safety signal.

\section{Outlook}
The framework is conceptual. We do not measure paired-error independence inside a deployed interactive workflow. Sites can run the audit only where they can adjudicate cases against a reference standard, and re-auditing after every change will be hardest where vendor updates are frequent. Current reporting standards for clinical AI do not, to our knowledge, require same-case error overlap. Even where mechanisms diverge, both clinician and model may fail on ambiguous or atypical cases, leaving their errors partly correlated. In screening mammography, for example, lesions that are difficult to diagnose by radiologists are also difficult for AI -- a partly shared blind spot rather than independent failure~\citep{r24}. We give a fuller treatment of the framework, the audit, and the governance model in a preprint~\citep{r25}, \href{https://doi.org/10.2196/preprints.107982}{doi:10.2196/preprints.107982}. The paired-error audit is one concrete step toward a standardized complementarity metric, and we see a prospective comparison against uniform or accuracy-triggered review as the natural next study.

\appendix
\section*{Revisions summary}
\addcontentsline{toc}{section}{Revisions summary}
In this camera-ready version we respond to the reviews. In Section~\ref{sec:framework} we now state what three closely related groups have and have not done, and cite two further papers on human-AI complementarity and on governance frameworks for clinical AI. In Section~\ref{sec:audit} we add a worked example covering threshold selection, a separate ceiling for serious-harm cases, the reading of an underpowered result, the pass rate at two sample sizes, and how a site resolves a clinician-model disagreement. In Section~\ref{sec:governance} we walk through the six stages of Figure~\ref{fig:governance} in order and give each stage an action and a decision rule. We cite the preprint with its DOI, make the caption of Figure~\ref{fig:governance} name error independence rather than vulnerabilities as the property preserved, and set the bibliography at full size. We keep Figure~\ref{fig:venn} as a schematic, and we now say in its caption that the grouping is neither exhaustive nor validated. To our knowledge, no one has published the four-cell paired-error breakdown for unaided clinicians and a locked LLM on a deploying site's own cases, and producing one is the empirical work we propose.


\begin{thebibliography}{27}
\bibitem[Newman-Toker et al.(2024)]{r1} Newman-Toker DE, et al. Burden of serious harms from diagnostic error in the USA. \textit{BMJ Qual Saf}. 2024;33(2):109-120.
\bibitem[Sherbino et al.(2014)]{r2} Sherbino J, et al. Ineffectiveness of cognitive forcing strategies to reduce biases in diagnostic reasoning: a controlled trial. \textit{CJEM}. 2014;16(1):34-40.
\bibitem[WHO(2024)]{r3} World Health Organization. \textit{Ethics and governance of artificial intelligence for health: Guidance on large multi-modal models}. Geneva: World Health Organization; 2024.
\bibitem[FDA(2025)]{r4} U.S. Food and Drug Administration. \textit{Artificial Intelligence-Enabled Device Software Functions: Lifecycle Management and Marketing Submission Recommendations}. Draft guidance for industry and Food and Drug Administration staff. Issued January 7, 2025.
\bibitem[Bansal et al.(2021)]{r5} Bansal G, et al. Does the whole exceed its parts? The effect of AI explanations on complementary team performance. \textit{Proc 2021 CHI Conf Hum Factors Comput Syst}. 2021:1-16.
\bibitem[Tawfik et al.(2018)]{r6} Tawfik DS, et al. Physician burnout, well-being, and work unit safety grades in relationship to reported medical errors. \textit{Mayo Clin Proc}. 2018;93(11):1571-1580.
\bibitem[Ly(2021)]{r7} Ly DP. The influence of the availability heuristic on physicians in the emergency department. \textit{Ann Emerg Med}. 2021;78(5):650-657.
\bibitem[Bhattacharyya et al.(2023)]{r8} Bhattacharyya M, et al. High rates of fabricated and inaccurate references in ChatGPT-generated medical content. \textit{Cureus}. 2023;15(5):e39238.
\bibitem[Hager et al.(2024)]{r9} Hager P, et al. Evaluation and mitigation of the limitations of large language models in clinical decision-making. \textit{Nat Med}. 2024;30:2613-2622.
\bibitem[Chen et al.(2025)]{r10} Chen S, et al. When helpfulness backfires: LLMs and the risk of false medical information due to sycophantic behavior. \textit{NPJ Digit Med}. 2025;8(1):605.
\bibitem[Shyr et al.(2025)]{r11} Shyr C, et al. A statistical framework for evaluating the repeatability and reproducibility of large language models. [Preprint]. \textit{medRxiv}. 2025. doi:10.1101/2025.08.06.25333170.
\bibitem[Croskerry(2003)]{r12} Croskerry P. The importance of cognitive errors in diagnosis and strategies to minimize them. \textit{Acad Med}. 2003;78(8):775-780.
\bibitem[Ly et al.(2023)]{r13} Ly DP, Shekelle PG, Song Z. Evidence for anchoring bias during physician decision-making. \textit{JAMA Intern Med}. 2023;183(8):818-823.
\bibitem[Ke et al.(2024)]{r14} Ke Y, et al. Mitigating cognitive biases in clinical decision-making through multi-agent conversations using large language models: simulation study. \textit{J Med Internet Res}. 2024;26:e59439.
\bibitem[Mahajan et al.(2025)]{r15} Mahajan A, et al. Cognitive bias in clinical large language models. \textit{NPJ Digit Med}. 2025;8(1):428.
\bibitem[Kuncheva \& Whitaker(2003)]{r16} Kuncheva LI, Whitaker CJ. Measures of diversity in classifier ensembles and their relationship with the ensemble accuracy. \textit{Mach Learn}. 2003;51(2):181-207.
\bibitem[Knight \& Leveson(1986)]{r17} Knight JC, Leveson NG. An experimental evaluation of the assumption of independence in multiversion programming. \textit{IEEE Trans Softw Eng}. 1986;SE-12(1):96-109.
\bibitem[Eckhardt \& Lee(1985)]{r18} Eckhardt DE, Lee LD. A theoretical basis for the analysis of multiversion software subject to coincident errors. \textit{IEEE Trans Softw Eng}. 1985;SE-11(12):1511-1517.
\bibitem[Z{\"o}ller et al.(2025)]{r19} Z{\"o}ller N, et al. Human-AI collectives most accurately diagnose clinical vignettes. \textit{Proc Natl Acad Sci U S A}. 2025;122(24):e2426153122.
\bibitem[Everett et al.(2026)]{r20} Everett SS, et al. From tool to teammate in a randomized controlled trial of clinician-AI collaborative workflows for diagnosis. \textit{NPJ Digit Med}. 2026;9(1):409.
\bibitem[Wang et al.(2026)]{r21} Wang G, et al. Human-large language model collaboration in clinical medicine: a systematic review and meta-analysis. \textit{NPJ Digit Med}. 2026;9(1):195.
\bibitem[Dvijotham et al.(2023)]{r22} Dvijotham KD, et al. Enhancing the reliability and accuracy of AI-enabled diagnosis via complementarity-driven deferral to clinicians. \textit{Nat Med}. 2023;29(7):1814-1820.
\bibitem[Jain et al.(2025)]{jain2025} Jain R, Bridgers S, Janzer L, Greig R, Teh TH, Mikulik V. Human-AI complementarity: a goal for amplified oversight. [Preprint]. \textit{arXiv}:2510.26518. 2025.
\bibitem[Li et al.(2025)]{li2026} Li J, Zhou ZC, Wang ZC, Lv H. Prioritizing human-AI collaboration in healthcare: the TRIAD framework for trustworthy governance, real-world, and integrated adaptive deployment. \textit{Mil Med Res}. 2025;12(1):97. doi:10.1186/s40779-026-00684-w.
\bibitem[Qazi et al.(2026)]{r23} Qazi IA, et al. Automation bias in large language model-assisted diagnostic reasoning among physicians trained in AI literacy: a randomized clinical trial. \textit{NEJM AI}. 2026;3(5). doi:10.1056/AIoa2501001.
\bibitem[Jiang et al.(2024)]{r24} Jiang Z, et al. AI for interpreting screening mammograms: implications for missed cancer in double reading practices and challenging-to-locate lesions. \textit{Sci Rep}. 2024;14(1):11893.
\bibitem[Amoei et al.(2026)]{r25} Amoei M, Fraser K, Poenaru D. Designing for Complementary Vulnerabilities: A Governance Framework for Deploying Large Language Models in Clinical Decision-Making. [Preprint]. \textit{JMIR Preprints}. 2026:107982. doi:10.2196/preprints.107982. \url{https://preprints.jmir.org/preprint/107982}
\end{thebibliography}
\end{document}